\documentclass[10pt, onecolumn]{article}
\usepackage[breaklinks=true]{hyperref}
\usepackage{amsmath}
\usepackage{authblk}
\usepackage{float}
\usepackage[margin=1in]{geometry}
\usepackage{graphicx}
\usepackage{tikz}
\usetikzlibrary{arrows,shapes, shapes.gates.logic.US}
\usepackage[utf8]{inputenc}
\usepackage[numbers]{natbib}
\usepackage{tabularx}
\usepackage{subcaption}
\usepackage{url}
\usepackage{xcolor}
\usepackage{listings}
\usepackage[all]{nowidow}

\definecolor{myBlue}{HTML}{1A73E8}
\definecolor{myRed}{HTML}{D93025}
\definecolor{myYellow}{HTML}{F9AB00}
\definecolor{myGreen}{HTML}{1E8E3E}

\newcolumntype{n}{X}
\newcolumntype{s}{>{\hsize=.35\hsize}X}

\restylefloat{table}

\title{A General Purpose Transpiler for Fully Homomorphic Encryption}

\author{Shruthi~Gorantala}
\author{Rob~Springer}
\author{Sean~Purser-Haskell}
\author{William~Lam}
\author{Royce~Wilson}
\author{Asra~Ali}
\author{Eric~P.~Astor}
\author{Itai~Zukerman}
\author{Sam~Ruth}
\author{Christoph~Dibak}
\author{Phillipp~Schoppmann}
\author{Sasha~Kulankhina}
\author{Alain~Forget}
\author{David~Marn}
\author{Cameron~Tew}
\author{Rafael~Misoczki}
\author{Bernat~Guillen}
\author{Xinyu~Ye}
\author{Dennis~Kraft}
\author{Damien~Desfontaines}
\author{Aishe~Krishnamurthy}
\author{Miguel~Guevara}
\author{Irippuge~Milinda~Perera}
\author{Yurii~Sushko}
\author{Bryant~Gipson}

\affil{fhe-open-source@google.com}

\date{June 14, 2021}

\begin{document}
\maketitle
\begin{abstract}
Fully homomorphic encryption (FHE) is an encryption scheme which enables computation on encrypted data without revealing the underlying data. While there have been many advances in the field of FHE, developing programs using FHE still requires expertise in cryptography. In this white paper, we present a fully homomorphic encryption transpiler that allows developers to convert high-level code (e.g., C++) that works on unencrypted data into high-level code that operates on encrypted data. Thus, our transpiler makes transformations possible on encrypted data. 

Our transpiler builds on Google's open-source XLS SDK~\cite{xls} and uses an off-the-shelf FHE library, TFHE~\cite{tfhe}, to perform low-level FHE operations. The transpiler design is modular, which means the underlying FHE library as well as the high-level input and output languages can vary. This modularity will help accelerate FHE research by providing an easy way to compare arbitrary programs in different FHE schemes side-by-side. We hope this lays the groundwork for eventual easy adoption of FHE by software developers. As a proof-of-concept, we are releasing an experimental transpiler~\cite{transpiler} as open-source~software.
\end{abstract}

\vfill

\section{Introduction}

As cloud computing services continue to see widespread adoption, it becomes increasingly important for service providers to guarantee the security and privacy of the data of their customers. Fully homomorphic encryption (FHE) is a cryptographic technique that provides strong security guarantees since the server only ever has access to encrypted data. FHE has not seen widespread use so far, for two main reasons: FHE is still too computationally expensive to be practical, and developing FHE applications requires extensive cryptographic expertise. Fortunately, over the last few years, FHE has become much less computationally intensive, due to significant progress in hardware acceleration, efficient optimizations, and low-level implementations. However, the widespread adoption of FHE still requires available tools that allow software developers without cryptography expertise to incorporate FHE into their applications. Our work attempts to bridge this gap. We have built and open-sourced a proof-of-concept general-purpose transpiler~\cite{transpiler} that can automatically convert a regular program that works on unencrypted data into one that performs the same operations on encrypted data.

In the remainder of this paper, the term \emph{plaintext} refers to unencrypted data, and \emph{ciphertext} refers to encrypted data.

\section{Background}

This section provides context and details on fully homomorphic encryption and transpilers for fully homomorphic encryption.

\subsection{Fully Homomorphic Encryption}

A \emph{homomorphic encryption scheme} is an encryption scheme in which some operations can be performed on both plaintext or ciphertext inputs and outputs. Homomorphic encryption has a few flavors including \emph{partially homomorphic encryption} (PHE), \emph{fully homomorphic encryption} (FHE), and \emph{somewhat homomorphic encryption} (SHE).

In PHE, only a subset of all possible computations can be performed on ciphertext without decryption. For example, in the additively homomorphic Paillier cryptosystem~\cite{paillier}, the product of two ciphertexts is equivalent to the sum of two plaintexts. In the multiplicatively homomorphic RSA cryptosystem~\cite{rivest}, the multiplication of two ciphertexts is equivalent to the multiplication of two plaintexts.

Often called the ``Holy Grail'' of cryptography~\cite{tourky}, FHE is an encryption mechanism that allows both addition and multiplication (and therefore any arbitrary computation) to be performed on encrypted data~\cite{rivest2}. Though first proposed in the 1970s~\cite{rivest2}, a theoretically feasible construction was not introduced until 2009 by Craig Gentry~\cite{gentry}.

A SHE scheme~\cite{he} allows both addition and multiplication to be performed but only for a few computations, after which the ciphertext loses too much integrity, and can no longer be correctly decrypted.

Most modern FHE schemes rely on the ``Learning with Errors'' (LWE) technique~\cite{pisa}, which relies on \emph{noise} being added to ciphertexts.  As long as the noise is sufficiently small, ciphertext can be decrypted to the correct message. During homomorphic operations, the noise in the ciphertext grows. While this effect is negligible during additions, multiplying two ciphertexts significantly increases the total amount of noise. As a result, only a fixed number of consecutive multiplications (a parameter called multiplicative depth) can be performed before decryption becomes impossible. This limitation can be circumvented using \emph{bootstrapping}, a technique that resets the noise level of a ciphertext to a fixed lower level by homomorphically evaluating the decryption circuit (i.e., the logic that converts ciphertext into plaintext) with an encrypted secret key as~input.

The first generation FHE schemes were based on the original Gentry scheme~\cite{gentry2} and were slow, often requiring 30 mins~\cite{viand} for a single multiplication. These first generation schemes convert an SHE scheme into an FHE scheme through bootstrapping.

Second generation schemes, such as BGV~\cite{brakerski} and BFV~\cite{fan}, use a technique called \emph{leveled homomorphic encryption}, which involves choosing sufficiently large parameters to allow the required computation without having to perform bootstrapping. They also introduced \emph{SIMD-style batching}~\cite{smart}, an optimization technique where many messages are packed into a single ciphertext, to reduce the overall latency overhead.

Third generation schemes based on the GSW scheme~\cite{gentry3} focus on \emph{fast bootstrapping}, which reduces the time spent on bootstrapping by several orders of magnitude. However, fast bootstrapping does not allow SIMD-style bootstrapping and batching, which presents a tradeoff between latency and throughput when compared to second generation schemes (i.e., BGV, BFV). The TFHE scheme~\cite{chillotti} (itself based on GSW) uses fast bootstrapping.

Developing software that uses FHE poses a unique challenge in \emph{parameter selection}. For a specific computation, one must choose the right parameter set that avoids bootstrapping, still provides security, and ensures that the size of the ciphertext remains manageable. It also involves selecting the proper encoding scheme for the right computations. For example, binary encoding is preferable for Boolean operations (i.e., TFHE), while arithmetic encoding is preferable for arithmetic computations (i.e., BGV and BFV).

\subsection{FHE Transpilers}

FHE is getting closer to becoming practical, but it still requires significant expertise to incorporate it in software development. Beyond the understanding of required cryptographic parameters, prior work mostly focused on supporting low-level programming primitives (e.g., arithmetic operators or Boolean logic), so developers need additional expertise in low-level software design to use these primitives and build more complex programs. FHE application development currently lacks a generic way for developers without cryptography expertise to write code without having to understand the underlying scheme. An FHE transpiler can bridge this gap.

A transpiler is a tool that converts one high-level language into another high-level language. An FHE transpiler takes code written in a high-level language (e.g., C++) and produces equivalent code capable of processing encrypted inputs.

As recently mentioned by Viand et al.~\cite{viand}, a missing component of the FHE story is a series of clean abstraction layers that separate business logic (i.e., what a developer is trying to achieve) from intermediate representation (i.e., how lower-level systems might reason or optimize around this) and optimized low-level implementation (i.e., what libraries and backends might be used to support these first two layers). Ideally, the top-most layer would accept high-level languages as input while allowing intermediate layers to remain expressive enough to represent gate operations and arithmetic operations in order to take advantage of various FHE schemes and translate among them~\cite{boura}, if needed. We believe that our transpiler helps fill this~gap.

\begin{figure}
    \centering
    \lstset {
        language=c++,
        basicstyle=\ttfamily\footnotesize,
        breaklines=true,
        commentstyle=\color{myYellow},
        frame=none,                  
        keywordstyle=\color{myBlue},
    }

    \newsavebox{\cppone}
    \begin{lrbox}{\cppone}
        \begin{lstlisting}
int sum(int a, int b) {
  return a + b;
}
        \end{lstlisting}
    \end{lrbox} 
    
    \newsavebox{\cpptwo}
    \begin{lrbox}{\cpptwo}
        \begin{lstlisting}
#include <tfhe.h>

// Full adder
void sum (LweSample* result,
    const LweSample* a,
    const LweSample* b,
    const int nb_bits,
    const TfheKeySet* bk) {
  LweSample* carry = new_ciphertext(bk->params);
  LweSample* temp = new_ciphertext(bk->params);
    
  // Initialize the carry to 0
  bootsCONSTANT(&carry, 0, bk);
    
  // Compute bit wise addition
  for (int i = 0; i < nb_bits; i++) {
    // Compute sum
    bootsXOR(&temp, &a[i], &b[i], bk);
    bootsXOR(&result[i], &temp, &carry, bk);
    
    // Compute carry
    bootsAND(&carry, &carry, &temp, bk);
    bootsAND(&temp, &a[i], &b[i], bk);
    bootsOR(&carry, &temp, &carry, bk);
  }
    
  delete_ciphertext(carry);
  delete_ciphertext(temp);
}
        \end{lstlisting}
    \end{lrbox}
    
    \tikzset{
        basic/.style = {draw, thin, fill = white, align = center, inner sep = 0.25cm}
    }
    \begin{tikzpicture}[auto]
        \node[basic] (code1) {\usebox{\cppone}};
        \node[basic, right of = code1, node distance = 8.5cm] (code2) {\usebox{\cpptwo}};
        
        \path [draw, thin, -latex] (code1) -- (code2);
    \end{tikzpicture}
    \caption{Example of a transpiled C++ program..}
    \label{fig:transcpp}
\end{figure}
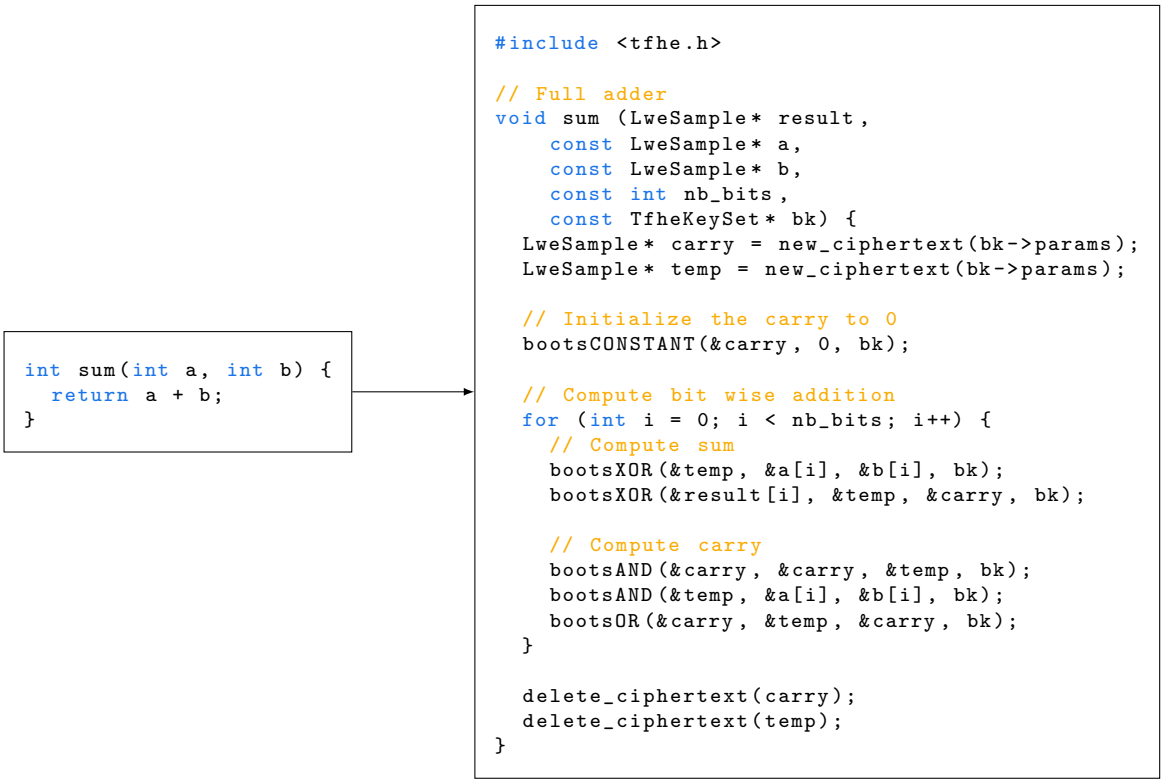

\section{Design}

Our general-purpose transpiler leverages features on the XLS toolchain and the gate level operations of TFHE to convert C++ programs into FHE-C++ programs. However, our transpiler design is modular, such that the backend FHE library, as well as the input and output languages, can be changed to suit different use cases.

\subsection{External Components}

Our transpiler uses two external components: XLS~\cite{xls} and TFHE~\cite{tfhe}.

\subsubsection{XLS}

XLS is a software development kit for hardware design. As part of its toolchain, it provides functionality to compile high-level hardware designs down to lower levels, and eventually to Verilog~\cite{verilog}. This compilation step introduces a flexible \emph{intermediate representation (XLS IR)} that allows for optimizations and other code transformations in the space between the high-level input language and the targeted output Verilog.

XLS IR is designed to describe and manipulate low-level operations (such as AND, OR, NOT, and other simple logical operations) of varying bit widths. Fortunately, this is exactly what is needed to translate higher-level language (e.g., C++) operations into lower-level Boolean operations (i.e., gates). Our transpiler uses XLS IR as the intermediate layer between the input C++ and the TFHE library. Optimizations (e.g., reducing gate count or eliminating unneeded bits) and transformations (e.g., \emph{Booleanification}, which is ``flattening'' an N-bit-wide operation to a series of 1-bit-wide operations) are performed on this intermediate~layer.

\subsubsection{TFHE}

TFHE is a fast fully homomorphic encryption library over torus, proposed by Chilloti et al.~\cite{chillotti2}. It is based on GSW and its ring variants. It significantly speeds up the bootstrapping operation (less than 0.1 sec~\cite{chillotti2}) and reduces bootstrapping key size, while preserving the same security levels as long as the T(R)LWE~\cite{chillotti2} problem remains intractable.

TFHE exposes an API for gate operations. In particular, it performs a bootstrap operation for every operation on the ciphertext. As the ciphertext refreshes itself after every gate operation, it allows for unlimited computations without noise management. With the operations provided by TFHE, it is possible to represent any computation as a composition of binary inputs and logical gates. This makes it ideal for the IR layer defined in the previous section to use TFHE for transpilation.

\subsection{Transpiler Invocation Stages}

We have designed a general-purpose transpiler for converting standard programs into FHE programs. As a proof-of-concept, we built a transpiler that uses TFHE to convert C++ programs into FHE-C++ programs, and published it as open-source software~\cite{transpiler}. However, our transpiler design is modular, such that the backend FHE library, as well as the input and output languages, can be changed to suit different use cases.

\begin{figure}
    \centering
    \tikzset{
        rect/.style        = {draw, fill = white, thin, align = center, rectangle, minimum width = 3.55cm, minimum height = 0.8cm},
        line/.style         = {draw, thin, -latex}
    }
    \begin{tikzpicture}[auto]
        \node (anchor1) {};
        \node[right of = anchor1, node distance = 6.0cm] (anchor2) {};
        \node[right of = anchor2, node distance = 6.0cm] (anchor3) {};
        
        \coordinate[above of = anchor3, node distance = 1.25cm] (edge1) {};
        \coordinate[below of = anchor3, node distance = 1.25cm] (edge2) {};
        
        \node[rect, above of = anchor1, node distance = 1.25cm] (r1) {C++};
        \node[rect, above of = anchor2, node distance = 1.25cm] (r2) {XLS IR};
        \node[rect, above of = anchor3, node distance = 0.0cm] (r3) {Optimized XLS IR};
        \node[rect, below of = anchor2, node distance = 1.25cm] (r4) {Booleanized XLS IR};
        \node[rect, below of = anchor1, node distance = 1.25cm] (r5) {FHE C++};
        
        \path [line] (r1) -- node {XLS[cc]} (r2);
        \path [line] (r2) -- node {Optimizer} (edge1) -- (r3);
        \path [line] (r3) -- (edge2) -- node {Booleanizer} (r4);
        \path [line] (r4) -- node[align = center] {FHE IR\\translation} (r5);
    \end{tikzpicture}
    \caption{High-level overview of the transpiler invocation process.}
    \label{fig:stages}
\end{figure}
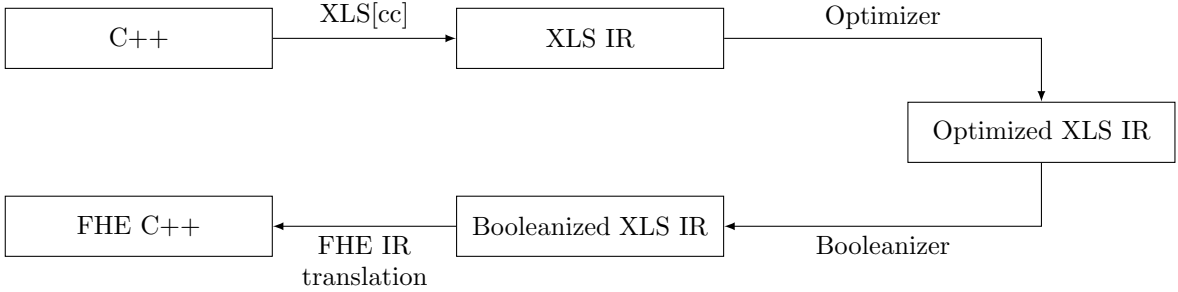

We now describe our transpiler by stepping through our initial open-source implementation. The sequential stages of a transpiler invocation are as follows:

\begin{itemize}
    \item \textbf{C++ frontend:} Converts C++ input into an intermediate representation (XLS IR).
    \item \textbf{Optimizer:} Simplifies the IR by reducing the number of operations, replacing operations with less-intensive equivalents, reducing bit widths, and other optimizations. This is part of the XLS toolchain.
    \item \textbf{Booleanifier:} Replaces all multi-bit compound operations with equivalent sequences of single-bit fundamental (e.g., AND, OR, NOT) operations. This is also part of the XLS toolchain. 
    \item \textbf{FHE IR Translator:} Constructs a C++ function invoking the TFHE library routine corresponding to each Boolean IR operation.
\end{itemize}

\subsubsection{XLS[cc] Stage}

We use XLS[cc]~\cite{xlscc} as our frontend, compiling C++ into XLS IR. 
  
\subsubsection{Optimizer Stage}

We leverage the XLS optimizer to apply various optimizations~\cite{xlsopt} to the IR produced by XLS[cc]. Our transpiler supports multiple optimization passes and the IR is Booleanified between optimization passes to take advantage of bit-level optimization opportunities.

\subsubsection{Booleanifier Stage}

The optimized XLS IR is translated into Boolean XLS IR, which only uses Boolean gates. This is required because TFHE only supports Boolean gates. This step is applied after each optimization pass if multiple optimization passes are requested.

\subsubsection{FHE IR Translation Stage}

The TFHE transpiler backend translates the Boolean XLS IR into C++ code that invokes TFHE for gate implementations. Our library also includes an FHE IR Interpreter which provides more flexibility in execution strategies, including multicore dispatch for improved performance.

This backend is deliberately constructed to be easily ported to any library that provides implementations for gates. Thus, other FHE libraries that provide the same gate-based interface can be used as drop-in replacements for TFHE. As an example, we implemented a version that uses native C++ boolean operations (without any FHE functionality), which proved useful for debugging during transpiler development.

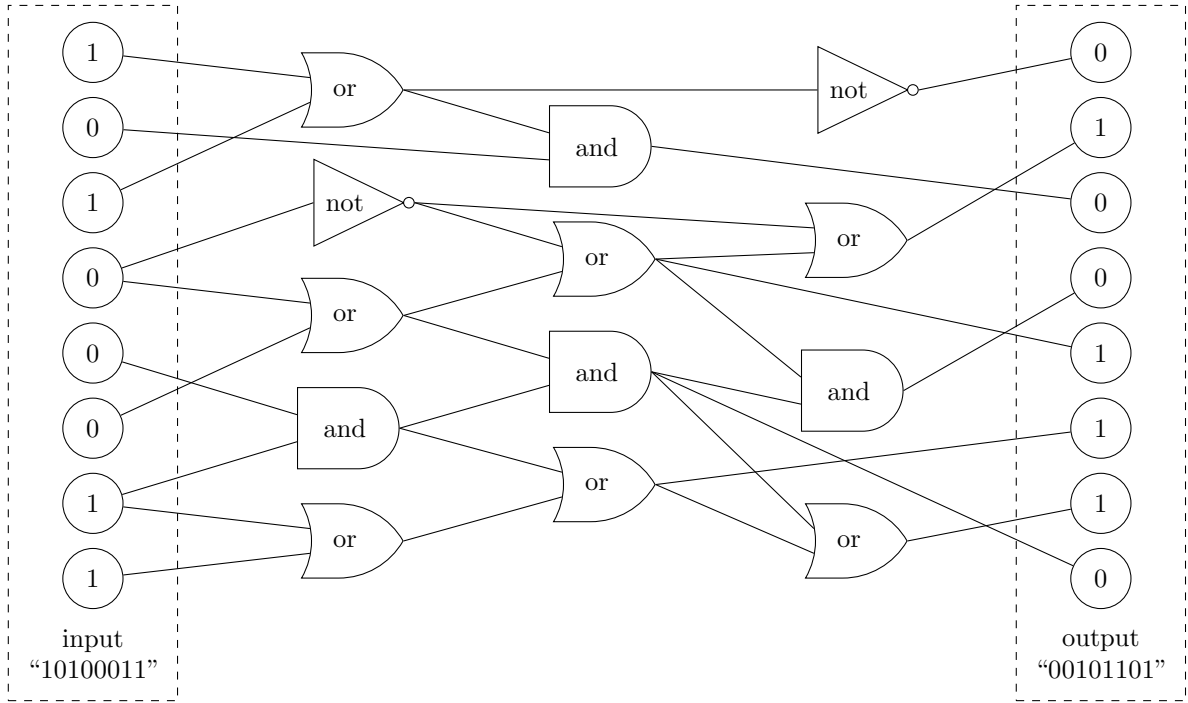
\begin{figure}
    \centering
    \tikzset{
        basic/.style        = {draw, fill = white, thin, align = center},
        bit/.style          = {basic, circle, minimum width = 0.8cm},
        and/.style          = {basic, and gate US, minimum width = 1.35cm, minimum height = 0.8cm},
        or/.style           = {basic, or gate US, minimum width = 1.35cm, minimum height = 0.8cm},
        not/.style          = {basic, not gate US, minimum width = 1.35cm, minimum height = 0.8cm},
        line/.style         = {draw, thin}
    }
    \begin{tikzpicture}[auto]
        \node (anchor1) {};
        \node[right of = anchor1, node distance = 3.35cm] (anchor2) {};
        \node[right of = anchor2, node distance = 3.35cm] (anchor3) {};
        \node[right of = anchor3, node distance = 3.35cm] (anchor4) {};
        \node[right of = anchor4, node distance = 3.35cm] (anchor5) {};
        
        \node[bit, above of = anchor1, node distance = 3.5cm] (i1) {1};
        \node[bit, above of = anchor1, node distance = 2.5cm] (i2) {0};
        \node[bit, above of = anchor1, node distance = 1.5cm] (i3) {1};
        \node[bit, above of = anchor1, node distance = 0.5cm] (i4) {0};
        \node[bit, below of = anchor1, node distance = 0.5cm] (i5) {0};
        \node[bit, below of = anchor1, node distance = 1.5cm] (i6) {0};
        \node[bit, below of = anchor1, node distance = 2.5cm] (i7) {1};
        \node[bit, below of = anchor1, node distance = 3.5cm] (i8) {1};
        
        \node[or, above of = anchor2, node distance = 3.0cm] (g11) {or};
        \node[not, above of = anchor2, node distance = 1.5cm] (g12) {not};
        \node[or, above of = anchor2, node distance = 0cm] (g13) {or};
        \node[and, below of = anchor2, node distance = 1.5cm] (g14) {and};
        \node[or, below of = anchor2, node distance = 3.0cm] (g15) {or};
        
        \node[and, above of = anchor3, node distance = 2.25cm] (g21) {and};
        \node[or, above of = anchor3, node distance = 0.75cm] (g22) {or};
        \node[and, below of = anchor3, node distance = 0.75cm] (g23) {and};
        \node[or, below of = anchor3, node distance = 2.25cm] (g24) {or};
        
        \node[not, above of = anchor4, node distance = 3.0cm] (g31) {not};
        \node[or, above of = anchor4, node distance = 1.0cm] (g32) {or};
        \node[and, below of = anchor4, node distance = 1.0cm] (g33) {and};
        \node[or, below of = anchor4, node distance = 3.0cm] (g34) {or};
        
        \node[bit, above of = anchor5, node distance = 3.5cm] (o1) {0};
        \node[bit, above of = anchor5, node distance = 2.5cm] (o2) {1};
        \node[bit, above of = anchor5, node distance = 1.5cm] (o3) {0};
        \node[bit, above of = anchor5, node distance = 0.5cm] (o4) {0};
        \node[bit, below of = anchor5, node distance = 0.5cm] (o5) {1};
        \node[bit, below of = anchor5, node distance = 1.5cm] (o6) {1};
        \node[bit, below of = anchor5, node distance = 2.5cm] (o7) {1};
        \node[bit, below of = anchor5, node distance = 3.5cm] (o8) {0};
        
        \node[below of = anchor1, node distance = 4.5cm, text width = 6.0em, align = center] (input) {input ``10100011''};
        \node[below of = anchor1, node distance = 0.5cm, draw, thin, minimum height = 9.25cm, minimum width = 2.25cm, dashed] (inputframe) {};
        
        \node[below of = anchor5, node distance = 4.5cm, text width = 6.0em, align = center] (output) {output ``00101101''};
        \node[below of = anchor5, node distance = 0.5cm, draw, thin, minimum height = 9.25cm, minimum width = 2.25cm, dashed] (outputframe) {};
        
        \path [line] (i1) -- (g11.input 1);
        \path [line] (i2) -- (g21.input 2);
        \path [line] (i3) -- (g11.input 2);
        \path [line] (i4) -- (g12.input);
        \path [line] (i4) -- (g13.input 1);
        \path [line] (i5) -- (g14.input 1);
        \path [line] (i6) -- (g13.input 2);
        \path [line] (i7) -- (g14.input 2);
        \path [line] (i7) -- (g15.input 1);
        \path [line] (i8) -- (g15.input 2);
        
        \path [line] (g11.output) -- (g21.input 1);
        \path [line] (g11.output) -- (g31.input);
        \path [line] (g12.output) -- (g22.input 1);
        \path [line] (g12.output) -- (g32.input 1);
        \path [line] (g13.output) -- (g22.input 2);
        \path [line] (g13.output) -- (g23.input 1);
        \path [line] (g14.output) -- (g23.input 2);
        \path [line] (g14.output) -- (g24.input 1);
        \path [line] (g15.output) -- (g24.input 2);
        
        \path [line] (g21.output) -- (o3);
        \path [line] (g22.output) -- (g32.input 2);
        \path [line] (g22.output) -- (g33.input 1);
        \path [line] (g22.output) -- (o5);
        \path [line] (g23.output) -- (g33.input 2);
        \path [line] (g23.output) -- (g34.input 1);
        \path [line] (g23.output) -- (o8);
        \path [line] (g24.output) -- (g34.input 2);
        \path [line] (g24.output) -- (o6);
        
        \path [line] (g31.output) -- (o1);
        \path [line] (g32.output) -- (o2);
        \path [line] (g33.output) -- (o4);
        \path [line] (g34.output) -- (o7);
    \end{tikzpicture}
    \caption{Visualization of a gate in XLS IR representation.}
    \label{fig:xlsgate}
\end{figure}

\subsection{Features}

\subsubsection{Modular Design}

Our transpiler design is modular in three ways:
\begin{itemize}
    \item The input code can be in any language that can be translated into XLS. This significantly reduces the burden of transpiling existing code, since it need not be written in a fixed supported language.
    \item The output FHE code can be in any language with an FHE library. This reduces the requirements for interacting with transpiled FHE code, since the FHE code can be transpiled into a language that interfaces well with the rest of the code providing the computational service.
    \item The underlying FHE backend can be any library that exposes gates as part of its API. Our library includes classes prefixed with \texttt{Encoded}, which can be reused to ease development. This can accelerate FHE research by providing an easy way to compare arbitrary programs in different FHE schemes side-by-side.
\end{itemize}

We hope our modular transpiler design will lay the groundwork towards more research breakthroughs and eventual easy adoption of FHE by software developers.

\subsubsection{TFHE Interface Utilities}

To interface with a FHE-C++ program (i.e., run it on some inputs and receive the output), developers need to write code that takes some input, sends the encrypted input to the generated FHE-C++ code, and allocates memory to receive the result. In the special case of a \emph{test bench} running the FHE-C++ program with known inputs, it also needs to encrypt the input and decrypt the result to verify it is as expected. 

We help simplify these operations, providing an API to encode/encrypt and decode/decrypt data while automatically managing memory allocation. This includes wrappers for several standard data types (e.g., integers, arrays of integers, and strings). Our transpiler also automatically generates wrappers with the same API for user-provided data types (e.g., classes and structs). More specifically, \texttt{FheValue} and other classes with the \texttt{Fhe} prefix~\cite{fhedata} provide a simple interface for converting between native plaintext data and encrypted \texttt{LweSample*} for a user that has the secret key, while also handling basic memory management. To help interface with other potential backends, including the unencrypted Boolean backend, we provide similar classes (\texttt{EncodedValue} and others with the \texttt{Encoded} prefix~\cite{fhebool}) that convert between native plaintext data and a plaintext-bits representation.

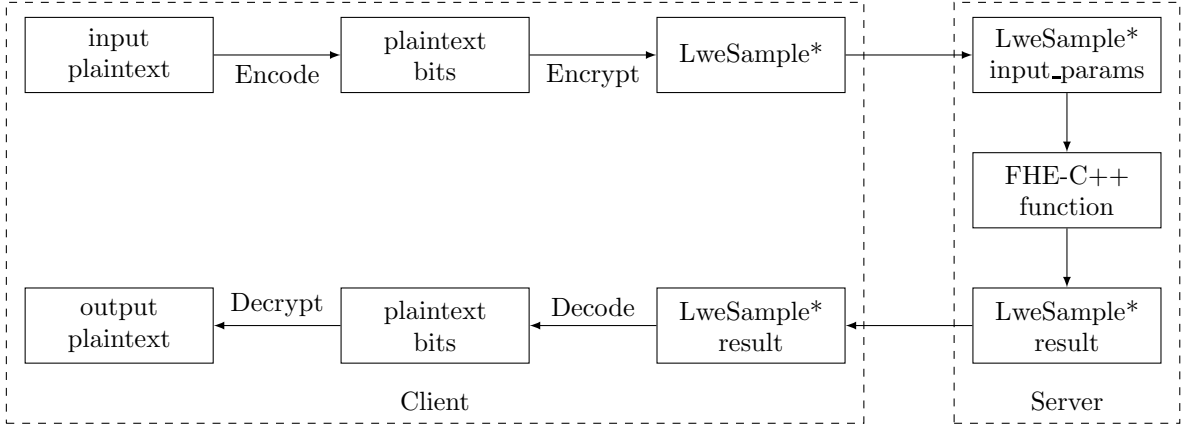
\begin{figure}
    \centering
    \tikzset{
        rect/.style        = {draw, fill = white, thin, align = center, rectangle, minimum width = 2.5cm, minimum height = 1.0cm},
        line/.style         = {draw, thin, -latex}
    }
    \begin{tikzpicture}[auto]
        \node (anchor1) {};
        \node[right of = anchor1, node distance = 4.2cm] (anchor2) {};
        \node[right of = anchor2, node distance = 4.2cm] (anchor3) {};
        \node[right of = anchor3, node distance = 4.2cm] (anchor4) {};
        
        \node[rect, above of = anchor1, node distance = 1.8cm] (r1) {input\\plaintext};
        \node[rect, above of = anchor2, node distance = 1.8cm] (r2) {plaintext\\bits};
        \node[rect, above of = anchor3, node distance = 1.8cm] (r3) {LweSample*};
        \node[rect, above of = anchor4, node distance = 1.8cm] (r4) {LweSample*\\input\_params};
        \node[rect, above of = anchor4, node distance = 0.0cm] (r5) {FHE-C++\\function};
        \node[rect, below of = anchor4, node distance = 1.8cm] (r6) {LweSample*\\result};
        \node[rect, below of = anchor3, node distance = 1.8cm] (r7) {LweSample*\\result};
        \node[rect, below of = anchor2, node distance = 1.8cm] (r8) {plaintext\\bits};
        \node[rect, below of = anchor1, node distance = 1.8cm] (r9) {output\\plaintext};
        
        \node[below of = anchor2, node distance = 2.8cm, text width = 6.0em, align = center] (client) {Client};
        \node[below of = anchor2, node distance = 0.3cm, draw, thin, minimum height = 5.6cm, minimum width = 11.4cm, dashed] (clientframe) {};
        
        \node[below of = anchor4, node distance = 2.8cm, text width = 6.0em, align = center] (server) {Server};
        \node[below of = anchor4, node distance = 0.3cm, draw, thin, minimum height = 5.6cm, minimum width = 3.0cm, dashed] (serverframe) {};
        
        \path [line] (r1) -- node[below] {Encode} (r2);
        \path [line] (r2) -- node[below] {Encrypt} (r3);
        \path [line] (r3) -- (r4);
        \path [line] (r4) -- (r5);
        \path [line] (r5) -- (r6);
        \path [line] (r6) -- (r7);
        \path [line] (r7) -- node[above] {Decode} (r8);
        \path [line] (r8) -- node[above] {Decrypt} (r9);
    \end{tikzpicture}
    \caption{Client server interaction.}
    \label{fig:clientserver}
\end{figure}

\subsubsection{Debugging Utilities}

One of the obstacles for FHE adoption is the inability to debug programs. To help with debugging, our open-source library includes an alternate backend that transpiles the input to a Booleanified C++ version (without any FHE features). This uses the same computation steps as the FHE version, but operates on plaintext bits. This is, of course, only intended to be used for debugging purposes, as it does not provide any security guarantees.

\section{How Best to Use the Transpiler}

Our transpiler is currently experimental. We do not recommend relying on it for production use cases. This open-source release is meant to showcase the feasibility of transforming general-purpose code (written by non-experts in cryptography) to FHE programs that work on encrypted data.

We see this as an important stepping stone in privacy-protecting technologies that can be iterated upon to accelerate research innovation and build feature-rich solutions. For example, as discussed in the Modular Design section above, the transpiler can support different input and output programming languages and can be used to compare different FHE schemes. This is possible because the XLS intermediate representation is independent of both the front-end language and the underlying FHE scheme. This also decouples FHE schemes from circuits, which allows researchers and experts in respective fields to make progress independently. 

\subsection{FHE Programming Restrictions}

Traditional imperative programming languages use data-dependent branching for control flow and optimizations. In FHE programming, computations must be \emph{data-independent}. The execution hardware doesn’t have access to the plaintext, so it cannot support branching. This implies a number of restrictions, as follows. 

Variable-length loops and arrays cannot be used, and must be replaced by fixed-sized arrays and loops with a fixed number of iterations. Early returns are not useful, since the entire function body must be evaluated regardless of the input. Recursion is not supported either, since it requires data-dependent termination. Pointers are not supported as they are data-dependent. Branch-and-bound optimizations are not possible because all branches must be executed.

\subsection{Threat Model}

Our transpiler generates code for a server to perform computations on encrypted text. It assumes an honest-but-curious adversary, whereby the server can be trusted to do the expected computation and knows the type of the data, but cannot be allowed to decrypt or view the data itself. The data type (e.g., int, string, class) is known to the server. The length of the data (e.g., the size of an input array) needs to be explicitly protected on the client-side by enforcing a fixed maximum input length (using padding).

The generated code does not protect against an adversarial server’s malicious manipulation of the data. This is inherent in all FHE schemes, given that the goal of FHE is to allow the server to compute functions on the data, but does not provide verification of what function was computed.

\subsection{Security Considerations}

The TFHE scheme~\cite{chillotti} bases its security on the torus variant of the (R)LWE problems. In cryptography, the reputation and credibility of a cryptosystem is typically established through work of cryptanalysis experts who have unsuccessfully tried to break such a scheme over multiple years. In contrast, the TFHE scheme is new and has not enjoyed as much public scrutiny as widely-deployed cryptographic primitives. The same paper that introduced TFHE~\cite{chillotti} presented a thorough cryptanalysis assessment, but the cryptography community may still eventually find more efficient algorithms that break the underlying premises of TFHE, which would impact the security estimates of the parameters used by the transpiler tool.

\section{Related Work}

Our FHE Transpiler's main novelty is that it generates debuggable high-level language output. Another primary contribution is the introduction of XLS as an expressive intermediate representation that is independent from both the front-end language and the underlying FHE scheme. This allows interoperability between languages and even FHE schemes. By decoupling the FHE scheme from the implementation circuits, researchers and experts in respective fields are free to make progress independently. Finally, it provides a framework to compare new FHE schemes and cryptographic optimizations uniformly on different programs.

However, the concept of automatically converting programming language source code into FHE-related implementations is not new. The Armadillo framework~\cite{carpov} (currently called the Cingulata toolchain), for instance, is also based on the TFHE scheme and converts C++ into Boolean circuits, performs gate operations, and converts the Boolean operations to FHE binary operations. In comparison to our work, Cingulata currently does not convert the source code back to any C++ source code. Furthermore, our transpiler supports higher-level operations such as SUM, MUL, and DIV in addition to gate operations. Converting the IR to FHE-C++ supports extensibility to a wider variety of FHE libraries and helps in debugging.

Another similar tool, also based on TFHE, is Encrypt-Everything-Everywhere (E3)~\cite{chielle}. It enables FHE operations, but requires the program to be written with overloaded methods provided by the E3 library. It also requires the user to specify a configuration on the data types used. E3 converts the compiled netlist into C++ functions that can be called directly from the program. In comparison, our transpiler works on pre-existing programs in supported high-level languages, automatically generating the transpiled FHE-C++ and relevant libraries to encode the data types involved in the program (including structs and classes) and does not need an explicit configuration file.

A few other similar tools, based on different (non-TFHE) underlying FHE schemes, have been proposed in the literature. For example, ALCHEMY~\cite{alchemy}, Marble~\cite{viand2} and RAMPARTS~\cite{ramparts} are all FHE compilation tools based on the BGV or FV schemes, which are good for homomorphic arithmetic operations but suffer from inefficient bootstrapping operations. There is also a growing literature on the specific topic of building FHE compiler tools for specific workloads. For example, nGraph-HE~\cite{boemer}, SEALion~\cite{sealion}, and CHET~\cite{dathathri} all intend to produce efficient and FHE-friendly code for certain machine learning workloads.

\section{Future Directions}

A natural direction for future work on a generic FHE transpiler like ours is to improve execution times. We believe that the XLS intermediate representation can help with reasoning about, and implementing, various optimizations. For instance, if programmable bootstrapping~\cite{chillotti3} is supported in the underlying FHE scheme, this would allow optimizations using high-performance versions of arbitrary univariate functions.

The transpiler currently uses bitwise operations, so all arithmetic operations are converted into many single-bit Boolean gates, making them quite slow. MUL and DIV/MOD of two ciphertext values are especially expensive. Support for native arithmetic operations and SIMD-style packing with CHIMERA~\cite{boura} would help speed up arithmetic operations and improve throughput.

The interpreter in the FHE IR Translation stage provides more flexibility in execution strategies, including multicore dispatch for improved performance. We intend to add additional optimizations that will be especially useful for heterogeneous compute environments, such as mixed CPU/GPU execution.

Automatic parameter selection has remained a challenging problem in the field of FHE because of the unknown circuits for programs. As our transpiler generates circuits for arbitrary programs, it could be augmented for automatic parameter selection at transpilation time.

\bibliographystyle{unsrt}
\bibliography{references}
\end{document}